\documentclass[preprint,showpacs,preprintnumbers,amsmath,amssymb,superscriptaddress]{revtex4}
\usepackage{amsmath}
\usepackage{amsmath}
\usepackage{amssymb}
\usepackage{amssymb}
\usepackage{amsmath}
\usepackage{amssymb}
\usepackage{graphicx}
\usepackage{txfonts}
\usepackage{subfigure}
\usepackage{dcolumn}
\usepackage{bm}

\begin{document}

\title{Time-dependent energetic proton acceleration and scaling laws in ultra-intense laser pulses interactions with thin foils}

\author{Yongsheng Huang }
\email{hyc05@mails.tsinghua.edu.cn}
\author{Yuanjie Bi }
\affiliation{China Institute of Atomic Energy, Beijing 102413, China.}%
\affiliation{Department of Engineering Physics, Tsinghua University, Beijing 100084, China.}%
\author{Yijin Shi }
\author{Naiyan Wang }
\author{Xiuzhang Tang }
\affiliation{China Institute of Atomic Energy, Beijing 102413,
China.}
\author{Zhe Gao}
\affiliation{Department of Engineering Physics, Tsinghua University, Beijing 100084, China.}%



\date{\today}

\begin{abstract}
A two-phase model, where the plasma expansion is an isothermal one
when laser irradiates and a following adiabatic one after laser
ends, has been proposed to predict the maximum energy of the proton
beams induced in the ultra-intense laser-foil interactions. The
hot-electron recirculation in the ultra-intense laser-solid
interactions has been accounted in and described by the
time-dependent hot-electron density continuously in this model. The
dilution effect of electron density as electrons recirculate and
spread laterally has been considered. With our model, the scaling
laws of maximum ion energy have been achieved and the dependence of
the scaling coefficients on laser intensity, pulse duration and
target thickness have been obtained. Some interesting results have
been predicted: the adiabatic expansion is an important process of
the ion acceleration and cannot be neglected; the whole acceleration
time is about $10-20$ times of laser pulse duration; the larger the
laser intensity, the more sensitive the maximum ion energy to the
change of focus radius, and so on.
\end{abstract}

\maketitle

\section{\label{sec:level1}Introduction}
Proton acceleration mechanisms in ultra-intense laser pulses
interaction with thin solid targets attract more and more interest
nowadays \cite{Machnisms,M.Kaluza2004,Emmanuel d'Humieres2005}.
Various models \cite{M.Kaluza2004,Schreiber2006,J.Fuchs,Robson2007N}
have been presented to estimate the maximum energy of proton beams.
However, the models given by Wilks {\it et al. }, Kaluza {\it et
al.}, Schreiber {\it et al.} and Fuchs {\it et al.} are all based on
isothermal expansions of quasi-neutral plasmas \cite{P.Mora2}.
Robson {\it et al.} presented a two-phase temperature-varying model,
where the hot-electron temperature first increases linearly on the
pulse duration timescale and then decreases adiabatically with time
. However, in the pulse duration, does the hot-electron temperature
rise up linearly? That is still difficult to be validated. Generally
ones assume that, when an ultra-intense laser pulse interacts with a
solid target, the laser-produced fast electrons with a uniform
temperature, $k_BT_e$, determined by the laser ponderomotive
potential are instantly created in front of the target and propagate
through the target collisionlessly and then form a high energy
plasma at the rear of the target. When the laser pulse still exists,
the hot-electron temperature, $k_BT_e=m_ec^2(\gamma-1)$, is assumed
invariant due to a constant energy supply from the laser pulse,
where $\gamma=(1+I\lambda^2/1.37)^{0.5}$ is the relativistic factor,
$I$ is the laser intensity in $\mathrm{10^{18}W/cm^2}$, $\lambda$ is
the laser wave length in $\mathrm{\mu m}$, $m_e$ is the electron
mass and $t_l$ is the pulse duration. The plasma expansion is an
isothermal expansion. Therefore, a two-phase model different from
Robson {\it et al.} is proposed in this article, where the plasma
expansion is isothermal in the laser pulse duration and then the
hot-electron temperature decreases as $(t/t_l)^{-(1+1/\gamma)}$
\cite{P.Mora3}.

The electron density distribution satisfies Boltzmann relationship:
$n_e=n_{e0}\mathrm{exp}(e\phi/k_BT_e)$ and $n_{e0}$ stays a constant
and time-independent in the previous
models\cite{M.Kaluza2004,Schreiber2006,J.Fuchs,P.Mora2} without
hot-electron recirculation, where $e$ is the elementary charge and
$\phi$ is the electric potential. Therefore, with a little
adjustment of some parameters: the acceleration time
\cite{Schreiber2006,J.Fuchs} , the opening angle of electrons
\cite{M.Kaluza2004} and electron density, $n_{e0}$
\cite{M.Kaluza2004,Schreiber2006,J.Fuchs}, Mora's result can be used
to estimate the maximum energy of proton beams for thick targets,
where the influence of hot-electron recirculation on the ion
acceleration can be ignored. Although Robson {\it et al.} have
presented a two-phase model which is consistent with experiments,
the hot-electron recirculation is still ignored. However, Mackinnon
et al. \cite{A.J.Mackinnon2002} observed enhancement of proton
acceleration by hot-electron recirculation in thin foils whose
thickness is less than a critical value. In addition, Sentoku {\it
et al.} \cite{Y.Sentoku2003} predicted an equation to conclude the
influence of electron recirculation and proved the hot-electron
recirculation cannot be ignored in the laser-foil interactions,
although they didn't propose a clear description of electron
recirculation and their physical picture is too simple and not
clear. The assumption: the maximum hot-electron density for a thin
foil is a constant and N times of the value for a critical target
thickness in Sentoku {\it et al.}'s model, is rough and
unreasonable. Because there are n times of the electron
recirculation they happen one after the other and the electron
density can't jump to $n$ times of the initial density. After that,
Huang {\it et al.}\cite{Huang2007re} presented a step model to
describe the influence of the hot-electron recirculation on the
laser-ion acceleration. In the step model, the hot-electron density
rises step by step with isothermal plasma expansions. In fact, the
electron density should rise continuously and then decrease to zero
as the time tends to infinite. Therefore, the time-dependent
hot-electron density and the electric field are necessary for the
description of the hot-electron recirculation and the whole process
of the ion acceleration. The dilution effect of the electron density
as the electrons circulate and spread laterally should be considered
but not accounted in the previous models
\cite{Huang2007re,Y.Sentoku2003}.

In Sec. \ref{sec:level11}, a new two-phase model which contains
three-dimensional effect (the thickness effect,  the angular effect,
which are discussed in detail by Huang {\it et al.}
\cite{Huang2007re},  and the dilution effect as the electrons
circulate and spread laterally) and the hot-electron recirculation
is proposed, where the plasma expansion is isothermal in the pulse
duration and then adiabatic. The main processes of our model are
two: first, combining the Mora's result in ref. \cite{P.Mora2} and
the increase of the electron density in the pulse duration, with the
assumption: the hot-electron temperature is a constant, the
dependence hot-electron density, the electric field and the ion
velocity on the time are obtained; second, with the assumption of an
adiabatic expansion, the dependence of the temperature of hot
electrons on time as proposed by Mora\cite{P.Mora3} has been used
and then the maximum ion velocity is obtained easily. A most
significant progress of our model is that: the time-dependent
electric field and hot-electron density can be given easily by
solving two nonlinear equations. As a result of the model, the
duration of the time-dependent electric field at the ion front is
approximately one to two times of the main laser pulse duration
which is consistent with the result presented by d'Humieres,
Lefebvre, Gremillet, and Malka in \cite{Emmanuel d'Humieres2005}
using the particle-in-cell (PIC) simulation. The whole acceleration
time is about $10-20$ times of the laser pulse duration. And we also
proofed that the adiabatic expansion is an important process for the
ion acceleration and cannot be neglected. Our model can be used in
the same application content as Robson's model: the target normal
sheath acceleration of ions, however, from the above discussions, it
is more reasonable and easily to use than their's.

In Sec. \ref{sec:level12}, with a proper laser absorption efficiency
 for thick targets, our two-phase model has been compared with experiments and they
are consistent as shown in Table \ref{tab:table1}. The laser
absorption stays constant with the target thickness for thick
target. The laser absorption efficiency for the target of arbitrary
thickness has been calculated by particle-in-cell (PIC) simulations
\cite{Emmanuel d'Humieres2005}, although there is no analytic result
of that. With the laser absorption efficiency of $40\%$ for the
target of $3\mathrm{\mu m}$ given by the result of PIC simulations,
the comparison between our model and the experimental result is
shown in Table \ref{tab:table1}. If the laser absorption is known,
for the target of arbitrary thickness, the maximum energy of proton
beams and time-dependent electric field and electron density can all
be obtained using our model.

In Sec. \ref{sec:level1d}, with our two-phase model, the scaling law
of maximum ion energy with respect to laser intensity for a series
of constant pulse duration has been given and discussed as shown in
Figures \ref{fig:parameters}. The dependence of maximum ion energy
on target thickness, focus radius, laser pulse duration have been
obtained. With the scaling law, some interesting results have been
obtained and discussed in detail. Also in this section, the limits
of our model have been discussed.

\section{\label{sec:level11}Time-Dependent Energy proton Acceleration}

For convenience, the physical parameters: the time, $t$, the ion
position, $l$, the ion velocity, $v$, the electron field, $E$, the
hot-electron density, $n$, and the light speed, $c$, are normalized
as Equation $(1)$ in \cite{Huang2007re}. Then the normalized
parameters are : $\tau$, $\hat{l}$, $u$, $\hat{E}$, $\hat{n}$,
$\hat{c}$ as shown by Equation $(1)$ in \cite{Huang2007re}.

When an ultra-intense laser pulse interacts with a solid target, the
laser-produced fast electrons with a uniform temperature, $k_BT_e$,
determined by the laser ponderomotive potential are instantly
created in front of the target and propagate through the target and
then form a high energy plasma at the rear of the target. Here, it
is assumed that the hot electron transport is collisionless, which
is true for high energetic electrons, thin foils or the atomic
number of the materials of the target is low. Hot electrons at the
rear of the target can be considered to be reflected by sheath field
at the ion front \cite{J.Santos2002,P.Mora2} and come back to the
front of the target, because the field there is strongest. Once hot
electrons are created, they will bounce between the ion front before
the target and the ion front at the rear side. Since we consider the
electron motion is collisionless, the bounce of hot electrons will
last in the whole time of the plasma expansion. When the hot
electrons propagate through the target, the electron beam can be
assumed to be in equilibrium.
\subsection{\label{sec:level2}Isothermal Expansion}
The hot-electron speed used is the light speed $c$. Here the choice
of $t=0$ is the same as that in the step model given by Huang and
co-workers in \cite{Huang2007re}. For simplicity in the $-L/c\leq
t\leq t_l-L/c$, where $L$ is the target thickness, the laser
intensity is assumed to be a constant, therefore, the hot-electron
temperature, $k_BT_e=m_ec^2(\gamma-1)$, is invariant. The plasma
expansion is an isothermal expansion.

The fast-electron density is a function of the parameters: the
acceleration time, $\tau$, the target thickness, $L$, laser
intensity, $I$, laser focus radius, $r_L$, the laser absorption
efficiency, $\eta$, the incidence angle of the laser pulse,
$\theta_{in}$, the half-opening angle of fast electrons, $\theta_e$.
The time-dependent electron density is assumed:
\begin{equation}\label{eq:neass}
\begin{array}{c}
n_{e}(\tau, L, I, r_L, \eta,\theta_{in}, \theta_{e})=N(\tau, L)n_{e0}(L, I, r_L, \eta, \theta_{in}, \theta_{e}),\\
N(\tau,L)=1, \tau=\tau_1=\tau_L,
\end{array}
\end{equation}
where $\tau_1$ is the time when the zeroth hot-electron
recirculation ends and hot electrons go forth to reach the rear of
the target the second time, $\tau_L=2\hat{L}/c/\sqrt{2e}$, here $e$
denotes the numerical constant 2.71828.... $n_{e0}(L, I, r_L, \eta,
\theta_{in}, \theta_{e})$ is the hot-electron density when hot
electrons return back from the ion front before the target and go
forth to reach the rear of the target the second time and $N(\tau,
L)$ describes the increase of the maximum electron density due to
electron recirculation and the electron generation by the
laser-plasma interactions at the front of the target.

Using Eq. (2) in \cite{M.Kaluza2004}, since the total number of hot
electrons that propagate through the target at $t=t_L=2L/c$,
$N_e=\eta(L)E_l/(k_BT_e)$ for $t_l\leq t_L$ and
$N_e=\eta(L)E_lt_L/(k_BT_et_l)$ for $t_l\geq t_L$, $n_{e0}$ in Eq.
(\ref{eq:neass}) can be estimated by:
\begin{equation}\label{eq:ne0Ll}
n_{e0}=
\frac{4.077\eta(L)I_{\mathrm{10^{18}W/cm^2}}}{(\gamma-1)(1+(L^{*}/r_L)
\rm{tan}(\theta_{e}))^2}, t_l\geq t_L\;
\end{equation}
\begin{equation}\label{eq:ne0Lg}
n_{e0}=\frac{4.077\eta(L)I_{\mathrm{10^{18}W/cm^2}}t_l}{(\gamma-1)(1+(L^{*}/r_L)
\rm{tan}(\theta_{e}))^2 t_L}, t_l\leq t_L,
\end{equation}
where $r_L$ is the laser pulse focus radius,
$L^{*}=L/cos(\theta_{in})$ is the efficient target thickness,
$\theta_{in}$ is the incidence angle of the laser pulse and
$\theta_{e}\approx17^o$ is half-opening angle of the superathermal
electrons which was measured by Santos {\it et al.}
\cite{J.Santos2002}. With Eq. (\ref{eq:ne0Ll}) and Eq.
(\ref{eq:ne0Lg}), the three-dimensional effect has been accounted in
through the considering of the half-opening angle of electrons,
$\theta_{e}\approx17^o$. Note that the right side of Eq.
(\ref{eq:ne0Lg}) has a factor, $t_l/t_L$, which is not in the right
side of Eq. (\ref{eq:ne0Ll}). For $t_l\leq t_L$,
$N_e=\eta(L)E_l/(k_BT_e)$, where $E_l$ is the energy of laser pulse.
However, for $t_l\geq t_L$, at $t=t_L$, hot electrons are still
being generated by the laser-plasma interactions at the front of the
target, and the number of hot electrons which propagate through the
target is a part of the total number,
$N_e=\eta(L)E_lt_L/(k_BT_et_l)$. Therefore, Equations
(\ref{eq:ne0Ll}) and (\ref{eq:ne0Lg}) are obtained. When $r_L\gg L$
and $tan(\theta_{in})\ll 1$, $(1+(L^{*}/r_L)
tan(\theta_{in}))^2\approx 1$, the angular effect can be neglected.
Therefore, the influence of $\eta(\hat{L})$ and electron
recirculation become dominated for thin targets. For example, for
$I=3\times 10^{20}\mathrm{W/cm^2}$, $\lambda=1.053\mathrm{nm}$ and
$t_l=500\mathrm{fs}$ \cite{R A Snavely}, the temperature of hot
electrons is about $5.5 \mathrm{MeV}$. For $\eta(L)=50\%$, $r_L=5\mu
m$, with Eq. (\ref{eq:ne0Lg}), the electron density is about
$4.8\times 10^{19} \mathrm{cm^{-3}}$ and it is about the value
measured by x-ray in \cite{R A Snavely}. Since the laser pulse
duration is $t_l$, with similar discussion in \cite{Huang2007re},
the critical target thickness for the hot-electron recirculation is:
$ L_c=0.5ct_l.$.

With reference to the discussion and method given by Huang and
co-workers in \cite{Huang2007re}, the relationship between the ion
velocity at the ion front and the electron density can be described
by Equation (12) in \cite{Huang2007re}. With that equation, the ion
velocity is decided by $N(\tau)$. Therefore, the solution of
$N(\tau)$ is a key point. Although it has been given by Huang {\it
et al.} \cite{Huang2007re} with a simple model, it is rough for
three reasons: $(1)$ it is discrete, however the actual electron
density change continuously; $(2)$ the electron density deceases as
electrons recirculate and spread laterally and this dilute effect
was not counted in the Step Model\cite{Huang2007re}; $(3)$ the
turning point of hot electrons at front of the target is not static
but moving with the expansion of the plasma too. In this paper, a
more actual and valuable method will be proposed to calculate a
continuous solution of $N(\tau)$ as follows.

\begin{figure}
{
\includegraphics[width=0.45\textwidth]{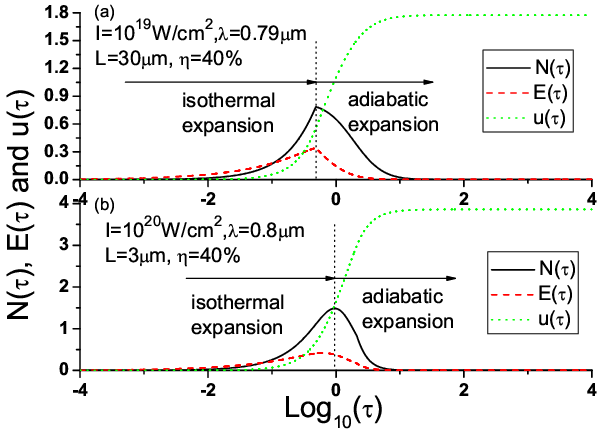}
}  \caption{\label{fig:nuE} (Color online) The time-dependent
hot-electron density, acceleration field and the speed of ions
versus $\tau=\omega_{pi}t/\sqrt{2e}$ given by the Time-Dependent
Target Normal Sheath Acceleration for $\theta_{in}=30^{o}$ (in Fig.
\ref{fig:nuE} (a)), $\theta_{in}=22^{o}$ (in Fig. \ref{fig:nuE}
(b)), $\theta_e=17^{o}$ and $\eta=40\%$. In Figure
\ref{fig:nuE}(a),the laser pulse parameters are $I=1.0\times
10^{19}\mathrm{W/cm^2}$, $\lambda=790\mathrm{nm}$,
$r_L=2.5\mathrm{\mu m}$, $L=30\mathrm{\mu m}$ and
$t_l=150\mathrm{fs}$. In Figure \ref{fig:nuE}(b), the laser pulse
parameters are $I=1.0\times 10^{20}\mathrm{W/cm^2}$,
$\lambda=800\mathrm{nm}$, $r_L=2.5\mathrm{\mu m}$, $L=3\mathrm{\mu
m}$ and $t_l=100\mathrm{fs}$. }
\end{figure}

Assuming that: the velocity at the ion front before the target is
the same as that at the rear of the target and the two turning
points for the electron recirculation are the ion front before the
target and at the rear respectively, if hot electrons satisfy
uniform distribution in the bulk from the ion front before the
target to the ion front at the rear, $N(\tau)$ is decided by
\begin{equation}\label{eq:Ntaultl}
N(\tau)=\frac{\int^{\tau-\hat{L}/\hat{c}}_{-\hat{L}/\hat{c}} f
d\tau}{\int^{\tau_L-\hat{L}/\hat{c}}_{-\hat{L}/\hat{c}} f
d\tau}\frac{\hat{L}+2\hat{l}(\tau_L)}{2\hat{l}(\tau)+\hat{L}}F_{\theta,3-D},
\tau\leq\tau_l,
\end{equation}
where $f$ represents the generation rate of hot electrons in the
interaction of laser pulses with the plasma at the front of the
target,
$F_{\theta,3-D}=\{\frac{f_{\theta}[\hat{L}+2\hat{l}(\tau_L)]+\hat{r}_L}{f_{\theta}[\hat{L}+2\hat{l}(\tau)]+\hat{r}_L}\}^2$
and $f_{\theta}=\tan(\theta)/\cos(\theta_{in})$. $f$ depends on the
absorption mechanisms of laser pulses and decides the density of hot
electrons. The factor, $F_{\theta,3-D}$, corresponds to the decrease
of on-axis density as hot electrons circulate and spread laterally
with an given opening angle, $\theta_{e}$. And also it reflects
three-dimensional effect on the ion acceleration. The special
integrating limits are because there is an interval before the
electrons generated by the laser pulse at the front of the target
come to the rear. With the assumption: $f=\bar{f}$ for
$\tau\in[0,\tau_l]$, Eq. (\ref{eq:Ntaultl}) can be simplified to be:
\begin{equation}\label{eq:Ntaultls}
N(\tau)=\frac{\tau}{\tau_L}\frac{\hat{L}+2\hat{l}(\tau_L)}{2\hat{l}(\tau)+\hat{L}}F_{\theta,3-D},
\tau\leq\tau_l,
\end{equation}

\subsection{\label{sec:level2}Adiabatic Expansion}
When $t\geq t_l$, the laser pulse has gone and the acceleration
field at the ion front decreases quickly for two reasons. First, the
temperature of hot electrons decreases with time as shown
by\cite{P.Mora3}:
\begin{equation}\label{eq:Tepro}
T_e\propto (\tau/\tau_l)^{-(1+1/\gamma)},
\end{equation}
For the nonrelativistic case, $\gamma=1$, with Eq. (\ref{eq:Tepro}),
$T_e\propto t^{-2}$ which is consistent with all the previous work
of adiabatic expansion into a vacuum\cite{Dorozhkina Baitin Kovalev,
M. A. True, G Manfredi} . For the ultra-relativistic case,
$\gamma\longrightarrow +\infty$, with Eq. (\ref{eq:Tepro}),
$T_e\propto t^{-1}$ which is the same with Mora's
results\cite{P.Mora3}. After the laser pulse vanishes, the ion front
does not stop and the electron bulk still increases. Therefore, the
electron density, $N(\tau)$, decreases as given by:
\begin{equation}\label{eq:Ntaugtls}
N(\tau)=\frac{\tau_l}{\tau_L}\frac{\hat{L}+2\hat{l}(\tau_L)}{2\hat{l}(\tau)+\hat{L}}F_{\theta,3-D},
\tau\geq\tau_l.
\end{equation}

Eq. (\ref{eq:Ntaultls}) and Eq. (\ref{eq:Ntaugtls}) are all
nonlinear differential equations and have no analytic solutions.
However, the numerical results can be obtained by computer with
iterative method. The initial $N(\tau)$ is given by the solution of
the Eq. (\ref{eq:Ntaultls}) and Eq. (\ref{eq:Ntaugtls}) in which
$F_{\theta, 3-D}\equiv1$. As an example, the solutions of a thin
foil and a thick solid target have been given by Figure
\ref{fig:nuE}. Figure \ref{fig:nuE}(a) corresponds a thick target of
$30 \mathrm{\mu m}$ and the hot electron recirculation can be
ignored. Figure \ref{fig:nuE}(b) corresponds a thin foil of
$3\mathrm{\mu m}$ and the maximum value of $N(\tau)$ is about $1.5$,
which is lower than that given by Huang and co-workers in
\cite{Huang2007re} and Sentoku et al. in \cite{Y.Sentoku2003} and
reflects the three-dimensional effect. From Fig. \ref{fig:nuE}, some
interesting results can be obtained:

$(1)$ The whole accelerate time of ions is about $10-20$ times of
the laser pulse duration. After that, the separating field is close
to zero and the acceleration ends.

$(2)$ The electron density and the electric field reach their
maximum value at the time $t=t_l$ as expected by our discussion and
the gain energy of ions in the process of the isothermal expansion
is approximately a quarter of the finally energy. Therefore, the
adiabatic expansion is also important for the ion acceleration
although the electron density and electric field decrease in this
process.

$(3)$ The influence of the hot-electron recirculation on the ion
acceleration for thin foils is obvious. The maximum ion energy for
thin foils is larger than that for thick targets.

With solutions of $N(\tau)$, the time-dependent electric field and
the ion velocity at the ion front can be obtained. Therefore, for
the target of arbitrary thickness, the maximum energy of proton
beams can be achieved if the laser absorption efficiency is known.

\section{\label{sec:level12}comparison with Experiments}

Our time-dependent model is compared with experiments, the results
are listed in Table \ref{tab:table1}.
\begin{table*}
\caption{\label{tab:table1}This is a comparison between our
two-phase model and some experiments for $\theta_{in}=30^{o}$ in
ref. \cite{M.Kaluza2004}, $\theta_{in}=22^{o}$ in ref.
\cite{A.J.Mackinnon2002} and $\theta_e=17^{o}$\cite{J.Santos2002}.}
\begin{ruledtabular}
\begin{tabular}{ccccccccc}
$I$  & $\lambda$  & $t_l$  & $r_L$  & $L$  & $\eta(L)$
& $n_{e0}(L)$ & $E_{max}$ from experiments & $E_{max}$ from our Model \\
($10^{18}\rm{W/cm^2}$) & ($\rm{\mu m}$) & ($\rm{fs}$) & ($\rm{\mu
m}$) & ($\rm{\mu m}$) & ($\%$) & ($\rm{10^{20}/cm^3}$) &
 or PIC ($\rm{MeV}$) & ($\rm{MeV}$)\\
\hline
10 & 0.79 & 150 & 2.5 & 30 & 40 & 0.33 & $1.2\pm 0.3$\cite{M.Kaluza2004} & 1.1\\
10 & 0.79 & 150 & 2.5 & 20 & 40 & 0.81 & $2.0\pm 0.3$\cite{M.Kaluza2004} & 2.0\\
13 & 0.79 & 150 & 2.5 & 30 & 40 & 0.36 & $1.5\pm 0.3$\cite{M.Kaluza2004} & 1.4\\
15 & 0.79 & 150 & 2.5 & 30 & 40 & 0.37 & $1.7\pm 0.3$\cite{M.Kaluza2004} & 1.6\\
100 & 0.8 & 100 & 2.5 & 3 & 40 & 14.2 & $22-24$\cite{A.J.Mackinnon2002} & 22.6\\
100 & 0.8 & 100 & 2.5 & 6 & 40 & 8.6 & $17-19$\cite{A.J.Mackinnon2002} & 17.3\\
100 & 0.8 & 100 & 2.5 & 10 & 40 & 5.13 & $11-17$\cite{A.J.Mackinnon2002} & 13.2\\
100 & 0.8 & 100 & 2.5 & 25 & 40 & 0.897 & $6-7$\cite{A.J.Mackinnon2002} & 5.0\\
\end{tabular}
\end{ruledtabular}
\end{table*}

For example, for $I=1\times10^{20}\mathrm{W/cm^2}$,
$\lambda=0.8\mathrm{nm}$ and $t_l=100\mathrm{fs}$, the critical
target thickness is about $15\mathrm{\mu m}$, according to Machinnon
{\it et al.} \cite{A.J.Mackinnon2002} , $E_{max}(L=30\rm{\mu
m})=6.2\mathrm{MeV}$. With the simulation results (Figure 12 in
\cite{Emmanuel d'Humieres2005}), the laser absorption stays constant
of about $35\%\to50\%$ with the target thickness for thick target,
$L\gtrapprox 1\mathrm{\mu m}$. The laser absorption changes with
target thickness, the contrast ratio between the main pulse and
prepulse and the prepulse duration \cite{M.Kaluza2004,Huang2007re}
for $L\lessapprox 1\mathrm{\mu m}$. For different target thickness,
the permeation of laser pulse is different. For different contrast
ratio and prepulse duration, the scale length of the preplasma is
different, which induces different laser absorption mechanism.
Therefore, the laser absorption efficiency, $\eta(L)$, is different
and difficult to be assured. Different $\eta(L)$ corresponds to
different electron density, $n_{e0}$. The plasma frequency and
acceleration parameters depend on $n_{e0}$. After all, the changing
law of $\eta(L)$ with $L$ for $L\leq 1\mathrm{\mu m}$ is quite
important for the proton acceleration and still a challenge. Without
$\eta(L)$, our model can not been compared with experiments for
$L\leq 1\mathrm{\mu m}$. However, the laser absorption efficiency
for the target of arbitrary thickness has been calculated by
particle-in-cell (PIC) simulations \cite{Emmanuel d'Humieres2005},
although there is no analytic result of that.

With the simulation results in ref. \cite{Emmanuel d'Humieres2005},
the small target thickness will lead to reduced absorption if the
target deconfines rapidly and becomes transparent before the end of
the laser pulse - but the characteristic velocity for this is the
sound speed, not the speed of light. If the critical thickness for
recirculation is $L_c$, the critical thickness for modified
absorption should be much smaller than $L_c$. Therefore, for thin
foils of the thickness, $L\geq 1\mathrm{\mu m}$, the laser
absorption efficiency keeps a constant about $35-50\%$ approximately
with the thickness, $L$.  For the target of $3\mathrm{\mu m}$,
$\eta\approx40\%$ and $\theta_{in}=17^{o}$, the maximum proton
energy is $22.6\mathrm{MeV}$ estimated by our model which is
consistent with the experimental data, $22-24\mathrm{MeV}$. The
time-dependent electron density, the acceleration field and the ion
speed are shown by Figure \ref{fig:nuE}(b). The hot-electron density
increases from $0$ and reaches the maximum value $1.5$ at the time
$t_l$. Therefore, for $L=3\mu m$ and $L_c=15\mu m$, the hot-electron
recirculation does exist and $N(\tau)$ is up to about $1.5$ but not
$5$ as shown by Sentoku {\it et al.} \cite{Y.Sentoku2003}. After
$t_l$, the electron density decreases quickly to half at about
$2.2t_l$. The duration of the hot-electron density and field are
$2-5 t_l$.

The maximum value of $N(\tau)$ is smaller than or equal to $1$ as
shown by Figure \ref{fig:nuE}. Figure \ref{fig:nuE} shows that: the
maximum $N(\tau)$ is about $0.79$ which is smaller than $1$,
therefore there is no hot-electron recirculation phenomena for
$L\geq L_c$; as the time goes to infinite, the velocity of protons
is finite and the maximum energy is about $1.1\rm{MeV}$ while the
experimental data is $1.2\pm 0.3\rm{MeV}$; the duration of the
hot-electron density is about $3.6t_l$. Therefore, the duration of
the field at the ion front is about $2t_l$, which is consistent with
the simulation result\cite{Emmanuel d'Humieres2005}.

\section{\label{sec:level1d}Scaling Law and Discussion}
The laser intensity in our model is assumed to be a constant value
in the pulse duration. Under this assumption and for a fixed laser
energy, the dependence of maximum ion energy on the laser pulse
duration is easy to be obtained. There is an optimum pulse duration
for the target normal sheath acceleration of ions if the laser
energy, focus radius and absorption efficiency sustain constants. It
is a conflict of large acceleration gradient and long efficient
acceleration time. For long pulse duration, the intensity will be
low and the acceleration field will be low. For a high intensity,
the efficient acceleration time will be short. Therefore, there is
an optimum pulse duration in the ion acceleration.

For different focus radius and target thickness, the dependence of
maximum ion energy, $E_{max,i}$, on pulse duration can be also
obtained easily with our model. The results may be different since
the plasma density changes with $r_L$ and $L$ as shown by Eq.
(\ref{eq:ne0Ll}) and Eq. (\ref{eq:ne0Lg}). The wave length will not
influence the dependence of $E_{max,i}$ on pulse duration.

\begin{figure}
{
\includegraphics[width=0.45\textwidth]{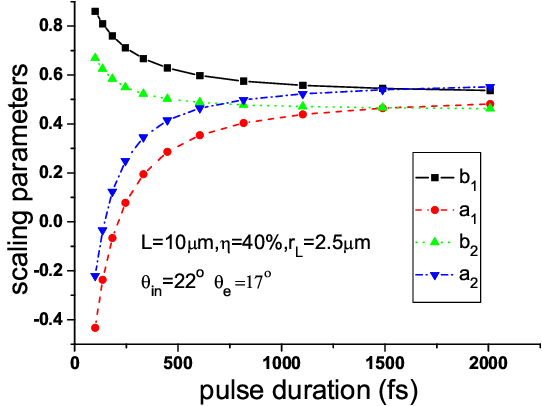}
}  \caption{\label{fig:parameters} (Color online) The coefficients
in the scaling law given by Eq. \ref{eq:scaling1} versus laser
intensity, $I$. Laser absorption efficiency is assumed $40\%$, since
the target thickness is large enough. Here, focus radius is
$2.5\mathrm{\mu m}$. }
\end{figure}

For a series of given pulse duration, the dependence of $E_{max,i}$
on laser intensity has also been obtained and the scaling law is
given by:
\begin{equation}\label{eq:scaling1}
E_{max,i}=\left\{ \begin{aligned}
         \mathrm{exp}(a_{1})(I_{10^{18}\mathrm{W/cm^2}}\lambda_{\mathrm{\mu
m}}^2)^{b_{1}}, I_{10^{18}\mathrm{W/cm^2}}\lambda_{\mathrm{\mu
m}}^2\lessapprox6.4, \\
\mathrm{exp}(a_{2})(I_{10^{18}\mathrm{W/cm^2}}\lambda_{\mathrm{\mu
m}}^2)^{b_{2}}, I_{10^{18}\mathrm{W/cm^2}}\lambda_{\mathrm{\mu
m}}^2\gtrapprox6.4,
\end{aligned} \right.
\end{equation}
where $a_1,a_2,b_1,b_2$ are all coefficients and shown by Fig.
\ref{fig:parameters}.

With Eq. (\ref{eq:scaling1}) and Fig. \ref{fig:parameters}, two
important results can be obtained. First, the scaling law is
different from the previous results $I^{1/2}$. The index $b_{1}$ and
$b_2$ depend on the laser pulse duration and decrease with pulse
duration. It shows the adiabatic expansion of plasmas is also a very
important acceleration process and should not be neglected, although
they\cite{J.Fuchs,M.Kaluza2004} can consist with experiments
considering the isothermal expansion only through adjusting of the
parameters: $\eta$, the acceleration time $t_{acc}$, the plasma
density $n_e$, the opening angle of electrons and so on. Second, the
influence of hot-electron recirculation on the ion acceleration can
be shown by Fig. \ref{fig:parameters} approximately.

\begin{figure}
{
\includegraphics[width=0.45\textwidth]{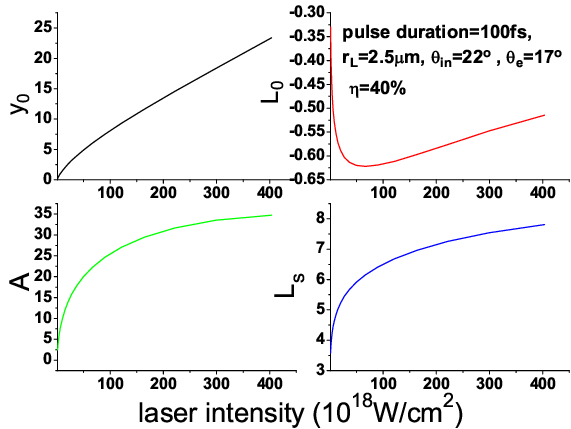}
}  \caption{\label{fig:scalingL} (Color online) The coefficients of
the scaling law of maximum ion energy with respect to target
thickness given by Eq. (\ref{eq:scalingL}) versus laser intensity
for $r_L=2.5\mathrm{\mu m}$, $\eta=40\%$, $t_l=100\mathrm{fs}$ and
$\lambda=0.8\mathrm{\mu m}$. }
\end{figure}

With our model, the dependence of maximum ion energy on target
thickness is given by:
\begin{equation}\label{eq:scalingL}
E_{max,i}=y_0+A\exp(-\frac{L-L_0}{L_s}),
\end{equation}
for $\eta=40\%$, $r_L=2.5\mathrm{\mu m}$, $\theta_{in}=22^o$,
$t_l=100\mathrm{fs}$ and $\theta_e=17^o$, where $y_0$, $A$, $L_0$,
$L_s$ are all coefficients. The dependence of them on laser
intensity have been shown in Fig. \ref{fig:scalingL}. Eq.
(\ref{eq:scalingL}) shows the maximum ion energy decreases with the
target thickness in the negative exponential form for a fixed laser
absorption and a given laser intensity. However, the laser
absorption tends to zero as $L\to 0$, therefore, the maximum ion
energy tends to zero in fact. From Fig. \ref{fig:scalingL}, maximum
ion energy increases with the laser intensity since the coefficients
$y_0, A$ all increase with the laser intensity.

\begin{figure}
{
\includegraphics[width=0.45\textwidth]{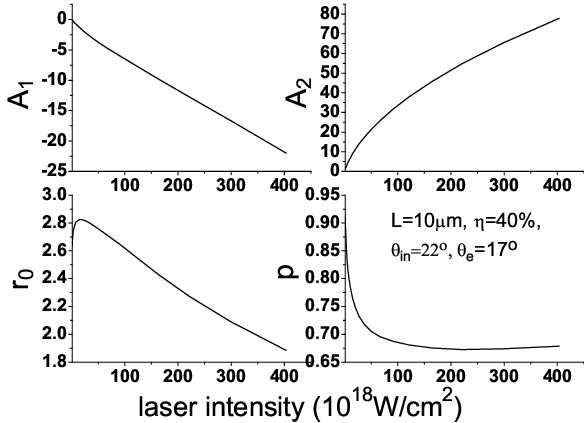}
}  \caption{\label{fig:scalingrL} (Color online) The coefficients of
the scaling law of maximum ion energy with respect to focus radius
given by Eq. (\ref{eq:rLscaling}) versus laser intensity for
$L=10\mathrm{\mu m}$, $\eta=40\%$, $t_l=100\mathrm{fs}$ and
$\lambda=0.8\mathrm{\mu m}$. }
\end{figure}

In fact, focus radius influences the electron density and then the
ion acceleration. For $L=10\mathrm{\mu m}, \eta=40\%,
\theta_{in}=22^o, \theta_e=17^o$, using our model, the effect of
focus radius satisfies:
\begin{equation}\label{eq:rLscaling}
E_{max,i}=A_2+\frac{A_1-A_2}{1+(r_L/r_0)^p},
\end{equation}
where $A_1$, $A_2$, $r_0$ and $p$ are all coefficients and change
with laser intensity and shown by Fig. \ref{fig:scalingrL}. With Eq.
(\ref{eq:rLscaling}), some interesting results can be achieved:

$(1)$ The smaller laser intensity, the larger the critical value
$r_0$ and the index $p$. Therefore, for $r_L\gg r_0$ and
$(r_L/r_0)^p\gg1$, $E_{max,i}\approx -(r_L/r_0)^{-p}$. The
derivation of $E_{max,i}$: $d E_{max,i}/d
(rL/r_0)=p(r_L/r_0)^{-p-1}$, reflects the rate of change of
$E_{max,i}$ to the focus radius. The rate of change is positive but
decreases with focus radius. For larger laser intensity, the larger
the rate of change.

$(2)$ Oppositely, for $r_L\ll r_0$ and $(r_L/r_0)^p\ll1$,
$E_{max,i}\approx constant$, which shows the influence of $r_L$ on
the maximum ion energy can be ignored in this case. For larger laser
intensity, the smaller critical focus radius. Therefore, the larger
laser intensity, the more sensitive the ion acceleration to the
change of focus radius.

The influence of opening angle of hot electrons on maximum ion
energy has been discussed in detail by Huang and co-workers in ref.
\cite{Huang2007re}. However, the dilution of the electron density as
electrons circulate and spread laterally was not contained there.
This effect is considered here with the factor, $F_{\theta, 3-D}$ in
Equations (\ref{eq:Ntaultl})-(\ref{eq:Ntaugtls}). Therefore, the
maximum value of electron density here is $1.5$ (in Fig.
\ref{fig:nuE}(b)) while it is about $5$ in ref. \cite{Huang2007re}
and ref. \cite{Y.Sentoku2003}. For the target of arbitrary
thickness, the maximum energy of ions heated by target normal sheath
acceleration (TNSA) can be obtained by this model if the absorption
efficiency of laser pulse is given.

Here we will discuss the limits of our model. The prepulse is not
considered in our model and the contrast is assumed large about
$10^8$ which can be achieved in lots of experiments. However, the
exist of a prepulse would generate a preplasma and the preplasma
size, the scaling length of the preplasma, will most influence the
mechanisms of laser absorption and then the temperature of hot
electrons. Different laser absorption mechanism results in different
generation rate of hot electrons, $f$, and different hot-electron
temperature, $T_e$. No matter what the mechanisms are, the
generation of hot electrons is cumulative and the assumption:
$f=\bar{f}$ causes little error relative to that caused by the
measurement in experiments. Whatever the temperature of hot
electrons is, our model is still in use with the actual temperature
instead of the value, $mc^2(\gamma-1)$. The laser intensity in our
model is assumed to be a constant value in the pulse duration. In
fact, the intensity is changing with time and the distribution is
about Gaussian distribution. However, the actual intensity
distribution with respect to time and position when laser pulse is
acting on a target is quite difficult to be measured in real time.
Since we do not consider the time distribution of laser intensity,
we can not give an estimation of the error. In the next paper, we
will consider that case.

\section{\label{sec:level1}Conclusion}
In conclusion, the time-dependent isothermal expansion and adiabatic
expansion for the target normal sheath proton acceleration is
discussed. A two-phase model and a new scaling law of the maximum
energy of proton beams have been proposed. The influence of the
hot-electron recirculation on the ion acceleration has been
accounted in. For $L\geq L_c$, the hot-electron recirculation can be
ignored. But for $L\leq L_c$, the hot-electron recirculation exists
and enhances as the target thickness decreases. The results given by
our model have been compared with experiments and shown in Table
\ref{tab:table1}. The dependence of maximum ion energy on target
thickness, focus radius, laser pulse duration have been obtained and
shown by equations (\ref{eq:scaling1}), (\ref{eq:scalingL}),
(\ref{eq:rLscaling}), and Figures \ref{fig:parameters},
\ref{fig:scalingL}, \ref{fig:scalingrL} and so on. At last, the
application and limits of our model has been discussed.

However, for thin foils, the laser absorption efficiency is an
important parameter for our model and is still a challenge for this
problem. The generation rate of hot electrons in the interaction of
laser pulses with the plasma at the front of the target, $f$, is
also a challenge. An interesting work that may be considered nest is
the time-dependent laser pulse intensity in order to optimize our
model further more.

This work was supported by the Key Project of Chinese National
Programs for Fundamental Research (973 Program) under contract No.
$2006CB806004$ and
 the Chinese National Natural Science Foundation under contract No.
$10334110$.

\end{document}